\documentclass[apj]{emulateapj-rtx4}

\usepackage{amssymb}
\usepackage{amsmath}
\usepackage{epsfig}
\usepackage{natbib}

\begin{document}

%\title{Afterglow Population Studies from Swift Follow-up of Fermi-LAT GRBs}
\title{Fermi and Swift Gamma-Ray Burst Afterglow Population Studies}

\author{J. L. Racusin\altaffilmark{1}, S. R. Oates\altaffilmark{2},
  P. Schady\altaffilmark{3}, D. N. Burrows\altaffilmark{4}, 
M. de Pasquale\altaffilmark{2}, D. Donato\altaffilmark{1,5,6},
N. Gehrels\altaffilmark{1},
S. Koch\altaffilmark{4}, J. McEnery\altaffilmark{1},
T. Piran\altaffilmark{7}, P. Roming\altaffilmark{4,8},
T. Sakamoto\altaffilmark{1}, C. Swenson\altaffilmark{4},
E. Troja\altaffilmark{1,9}, V. Vasileiou\altaffilmark{1,10,11}, 
F. Virgili \altaffilmark{12}, D. Wanderman\altaffilmark{7}, 
B. Zhang\altaffilmark{12}}
\altaffiltext{1}{NASA, Goddard Space Flight Center, Code 661, Greenbelt, MD, USA}
\altaffiltext{2}{Mullard Space Sciences Laboratory, University College London, Surrey, UK}
\altaffiltext{3}{Max-Planck-Institut fur extraterrestrische Physik,
  Garching, Germany}
\altaffiltext{4}{The Pennsylvania State University, University Park,
  PA, USA}
\altaffiltext{5}{Department of Physics and Department of Astronomy,
  University of Maryland, College Park, MD 20742, USA}
\altaffiltext{6}{Center for Research and Exploration in Space Science and Technology and NASA Goddard Space Flight Center, Greenbelt, MD 20771, USA}
\altaffiltext{7}{The Racah Institute of Physics, Hebrew University,
  Jerusalem 91904, Israel}
\altaffiltext{8}{Southwest Research Institute, Department of Space
  Science,6220 Culebra Rd, San Antonio, TX 78238, USA}
\altaffiltext{9}{Oak Ridge Associate Universities}
\altaffiltext{10}{University of Maryland, Baltimore County, MD, USA}
\altaffiltext{11}{Laboratoire de Physique Th´eorique et Astroparticules, Universit´e Montpellier 2, CNRS/IN2P3, Montpellier, France}
\altaffiltext{12}{University of Las Vegas, Las Vegas, NV, USA}

\slugcomment{Accepted for publication in ApJ}

\begin{abstract}
The new and extreme population of GRBs detected by {\it Fermi}-LAT shows
several new features in high energy gamma-rays that are providing
interesting and unexpected clues into GRB prompt and afterglow
emission mechanisms. Over the last 6 years, it has been {\it Swift} that has
provided the robust dataset of UV/optical and X-ray afterglow
observations that opened many windows into components of GRB emission
structure.  The relationship between the LAT detected GRBs and the well
studied, fainter, less energetic GRBs detected by {\it Swift}-BAT is only
beginning to be explored by multi-wavelength studies. We explore the
large sample of GRBs detected by BAT only, BAT and {\it Fermi}-GBM, and GBM
and LAT, focusing on these samples separately in order to search for
statistically significant differences between the populations, using
only those GRBs with measured redshifts in order to physically
characterize these objects. We disentangle which differences are
instrumental selection effects versus intrinsic properties, in order
to better understand the nature of the special characteristics of the
LAT bursts. 
\end{abstract}


\section{Introduction}

The field of gamma-ray bursts (GRBs) is undergoing dramatic changes
for a second time within the past decade, as a new observational window has
opened with the launch and success of NASA's {\it Fermi} gamma-ray
space telescope.  While both NASA's {\it Swift}
gamma-ray burst 
explorer mission \citep{gehrels04} and {\it Fermi} are operating
simultaneously, we have the ability to potentially detect hundreds
of gamma-ray bursts per year ($\sim 1/3$ of which are triggered by
{\it Swift}).  This allows prompt observations in the $15-150$ keV
hard X-ray band with the Burst Alert Telescope (BAT,
\citealt{barthelmy05}) and rapid follow-up in the $0.3-10$ keV soft 
X-ray band with the X-Ray Telescope (XRT, \citealt{burrows05}) and the
UV/optical band  by the Ultraviolet Optical Telescope (UVOT,
\citealt{roming05}) on-board {\it Swift}.  There is $\sim 40\%$ overlap between BAT triggers
and triggers from {\it Fermi}'s Gamma-ray Burst Monitor (GBM,
\citealt{meegan09}) allowing for coverage from 10 keV to 30 MeV, and a
special subset detected up to 10s of GeV with {\it Fermi}'s Large
Area Telescope (LAT, \citealt{atwood09}).  This 
wide space-based spectral window is broadened further by ground
based optical, NIR, and radio follow-up observations. 

In the last 2 years, the addition of the 30 MeV to 100 GeV
window from {\it Fermi}-LAT has led to another theoretical crisis, as
we attempt to understand the origin and relationship between these new
observational components and the ones traditionally observed from GRBs
in the keV-MeV band.
Just as {\it Swift} challenged our theoretical models by
demonstrating that GRBs have 
complex behavior in the first few hours after the trigger \citep{nousek06},
{\it Fermi}-LAT is regularly observing a new set of high energy
components in a small very energetic subset of bursts
\citep{abdo090902b,abdo080825c,abdo080916c,abdo081024b,abdo090510}.  The
relationship between the $>100$ MeV emission and the well studied
keV-MeV components remains unclear
\citep{corsi09,corsi10,kumar09,razzaque09,zhang09,ghisellini10,peer10,piran10,toma10,wang10}.

The complicated {\it Fermi}-LAT prompt emission spectra do not show simply
the extension of the lower-energy Band function \citep{band93}, but
rather the joint GBM-LAT spectral fits can also show the presence of an
additional hard power-law that can be detected both above and below
the Band function \citep{abdo090902b,abdo090510} in some cases.  There were earlier 
indications of this additional spectral component in the EGRET
detected GRB 941017 \citep{gonzalez03}.  However, the
rarity of EGRET GRB detections left it unclear whether this was a common
high energy feature, or if special circumstances in that
GRB were responsible.  This
component is too shallow to be due to Synchrotron self-Compton (SSC)
as had been predicted extensively pre-{\it Fermi}
\citep{zhang01,guetta03,galli08,racusin08,band09}.  The spectral
behavior of the LAT bursts appears to rule out the theory that the
soft $\gamma$-rays are caused by a SSC or another Inverse Compton (IC)
component \citep{ando08,piran09}.

{\it Fermi}-LAT's $>100$ MeV temporal behavior is different from
the lower-energy counterparts observed from thousands of GRBs.  The LAT
emission often begins a few seconds later than the lower-energy prompt
emission, and sometimes lasts substantially longer (up to thousands of
seconds; \citealt{abdo090902b,abdo080916c,abdo090510}).  This so-called ``GeV extended 
emission'' and the extra spectral power-law component may be the same
component, but the statistics in the extended emission are limited and
detailed spectral fits are often not possible. 
%  However, in at least
%the case of GRB 090902B \citep{abdo090902b}, the spectral indices of
%the two components (extra spectral power law and temporal extended
%emission) are inconsistent.

Several groups \citep{ghisellini10,kumar09,depasquale10} suggest that
the high-energy 
extended emission is caused by the same forward shock 
mechanism (with special caveats for environmental density and magnetic
field strength) responsible for the well studied broadband afterglows
that have been observed for hundreds of other GRBs.  Alternatively,
\cite{zhang11,maxham11} suggest that the $>100$ MeV emission during
the prompt emission phase is of internal origin, and the later
extended emission is of external origin.  We can learn more
about the mysterious new LAT components by studying the GRBs from a
broadband perspective, for which the early broadband afterglow (specifically X-ray
and optical) behavior is well studied.   However, currently there is
only one case of simultaneous X-ray/optical/GeV emission in the
minutes after the GRB - the short hard GRB 090510 which was
simultaneously triggered upon the {\it Swift}-BAT and the {\it
  Fermi}-GBM and LAT \citep{depasquale10}.

Despite the lack of simultaneous LAT and lower energy observations in
long bursts, we
can still learn about the special nature of the LAT bursts by studying
their lower energy late-time afterglow observations.  In this paper,
we utilize the large database of {\it Swift} afterglow observations of
BAT discovered bursts, and compare them to the simultaneous BAT/GBM
triggers, and the {\it Swift} follow-up of the LAT/GBM detected
bursts, in order to learn about the properties of the different
populations of GRBs.  

This paper is organized as follows: in Section \ref{sec:sample} we
discuss the sample selection and data analysis, in Section
\ref{sec:results} we discuss the results and correlations apparent in
the different samples, in Section \ref{sec:disc} we discuss the physical implications of
our analysis, and in Section \ref{sec:conc} we conclude.

\section{Sample Selection and Analysis}\label{sec:sample}
In order to study the population differences between the BAT-triggered
sample, the GBM-triggered sample, and the LAT detected sample, we use
all GRBs from these samples with measured redshifts and well
constrained XRT and UVOT light curves (at least 4 light curve bins and
enough counts to construct and fit an X-ray spectrum).

As of December 2009, the {\it Swift} XRT and UVOT instruments have
observed afterglows of 439 bursts discovered by BAT, as well as
81 bursts discovered by other missions.
We now have enough detailed
observations of X-ray and UV/optical afterglows from {\it Swift} to study them as a
statistical sample, and to attempt to separate observational biases from
physical differences in GRB populations.  In this study, we included
afterglow observations of
all GRBs discovered by {\it Swift}-BAT between December 2004 and December
2009 (sample hereafter referred to as BAT), with measured redshifts in
the literature (e.g. \citealt{fynbo09}) and well-constrained light curves
and spectra.

Unfortunately, the positional errors provided by GBM are too large
(several degrees) to facilitate {\it Swift}
follow-up.  Therefore, the only afterglow observations we have of GBM triggered
bursts, are those that simultaneously triggered the {\it Swift}-BAT and meet
all of the same criteria as the BAT sample (sample
hereafter referred to as GBM).  We treat the GBM bursts separately from the
BAT sample, and do not include them in both samples.  The GBM
bursts have a much wider measured spectral range during the prompt
emission than the BAT bursts, and are therefore more likely to provide
an accurate measurement of 
$E_{peak}$ and the Band function parameters.  From an instrumental
perspective, BAT has a much better sensitivity than GBM.  Therefore,
the GBM bursts are biased towards higher fluence. 

We also include the LAT detected GBM triggered GRBs that have been
localized by XRT and UVOT in follow-up observations 
(sample hereafter referred to as LAT).  The LAT position
errors for those GRBs for which follow-up was initiated were $\sim
3-10$ arcminutes radius.  Those GRBs with initial position 
errors that were significantly larger than the XRT field of view (FoV) have not been
followed-up by {\it Swift}.  The LAT sample includes the one BAT/GBM/LAT
simultaneous trigger (GRB 090510), and the remaining LAT detected GRBs
that BAT did not observe (i.e. were outside the BAT FoV at trigger).  There has
been follow-up of 10 out of 24 LAT detected GRBs (as of 
March 2011) with position errors small enough to initiate Target of
Opportunity (ToO) observations.  However, (with the exception of the
joint BAT/GBM/LAT trigger) these  
follow-up observations began at a minimum of 12 hours after the
trigger, and in some cases, did not begin until 1-2 days
post-trigger.  Despite this impediment, 8 of 10 LAT GRBs were
detected by XRT, and 7 of 10 by UVOT (5 in u-band, see Section \ref{sec:uvot}).

The breakdown of GRBs in each of the XRT and UVOT
datasets for the BAT, GBM, and LAT samples are given in Table
\ref{tab:numbers}.  Individual bursts are included in only one of the
BAT, GBM, and LAT samples, depending on their detection by one, two,
or all three instruments.  In the following Sections, we describe the data
analysis of the follow-up X-ray and UV/optical observations, as well
as the methods and sources for obtaining $\gamma$-ray prompt emission
spectral fits and fluences.

\begin{deluxetable}{lcc}
\tablewidth{3in}
\tablecolumns{3}
\tablecaption{Sample Statistics \label{tab:numbers}}
\tablehead{
\colhead{Sample} & \colhead{XRT} & \colhead{UVOT}}
\startdata
BAT & 147 & 49 \\
GBM & 19 & 11 \\
LAT & 8 & 5
\enddata
\tablecomments{The number of GRBs in each sample that meet our
  selection criteria.  For a GRB to be
  included in the UVOT sample, observations in the u-band are needed
  for normalization.}
\end{deluxetable}

\subsection{X-ray} \label{sec:xray}
The X-ray light curves and spectral fits were obtained from the XRT team 
repository \citep{evans07,evans09}.  We fit and characterized all of
the light curves using the methods of \cite{racusin09}.  Each count
rate light
curve was fit with the best fitting model of either a power law,
broken power law, double broken power law, or triple broken power law,
after time periods of significant flaring were manually removed.

\begin{figure}[b]
  \begin{center}
    \includegraphics[scale=0.35,angle=90]{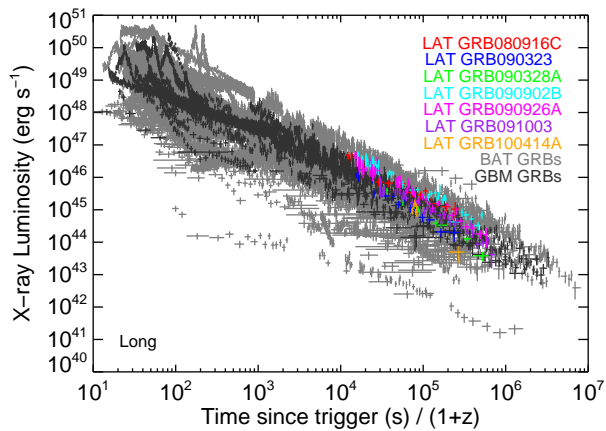}
    \includegraphics[scale=0.35,angle=90]{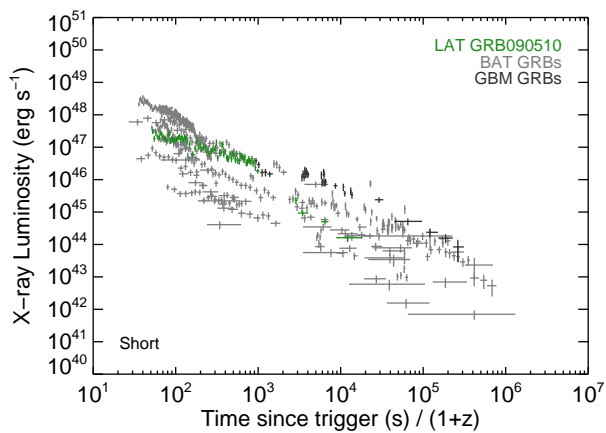}
  \end{center}
  \caption{The X-ray (0.3-10 keV) afterglow rest-frame luminosity light curves for
    all of the long ({\it top panel}) and short ({\it bottom panel})
    GRBs in our samples.  The BAT, GBM, and LAT GRBs are
    indicated by the different colors.  Note the clustering of the LAT
    light curves compared to the BAT and GBM samples in the top panel.
    \label{fig:xlc}}
\end{figure}

We convert the count-rate light curves to flux light curves based on a
single counts-to-flux conversion factor obtained from the Photon
Counting (PC) mode
spectral fit.  In order to physically characterize the afterglows,
taking advantage of the redshift information, we convert flux to
luminosity and apply a k-correction using the following formalism
from \cite{berger03}: 
\begin{equation}\label{eq:lx}
L_x(t)=4\pi D_L^2 F_x(t)(1+z)^{\alpha_x-\beta_x-1}
\end{equation}
where $L_x(t)$ is the $0.3-10~\textrm{keV}$ luminosity at time $t$
seconds after the trigger; $F_x$ is the $0.3-10~ 
\textrm{keV}$ flux at time $t$; $D_L$ is the luminosity
distance assuming cosmological parameters: $H_0=71~
\textrm{km}~\textrm{s}^{-1}~\textrm{Mpc}^{-1}$, $\Omega_m=0.27$, and
$\Omega_{\Lambda}=0.73$; $\alpha_x$
is the X-ray power-law temporal decay at time $t$; and $\beta_x$ is
the spectral energy index at time $t$.  The X-ray luminosity light
curves in rest frame time for both the long and short bursts, for all
three of our samples, are shown in Figure \ref{fig:xlc}.

The X-ray spectra were also taken from the XRT team repository.
\cite{evans09} describes how the spectra are extracted and fit to
absorbed power-laws, with two absorption components 
(Galactic and intrinsic at the GRB redshift).  The spectral power-law index
($\Gamma$) is converted to the energy index via $\beta=\Gamma-1$,
where $F_x=t^{-\alpha_x}\nu^{-\beta_x}$.

 \begin{figure}[b]
   \begin{center}
   \includegraphics[scale=0.35,angle=90]{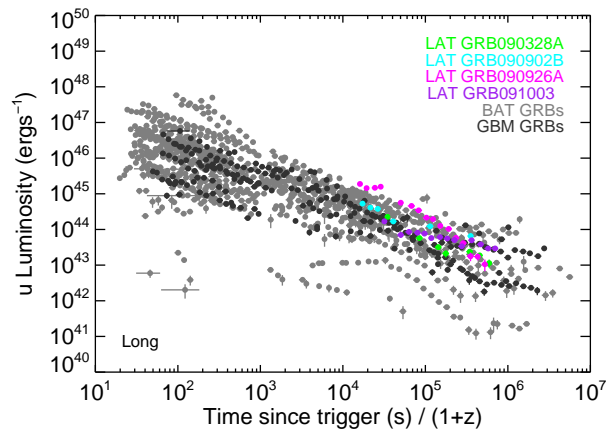}
   \includegraphics[scale=0.35,angle=90]{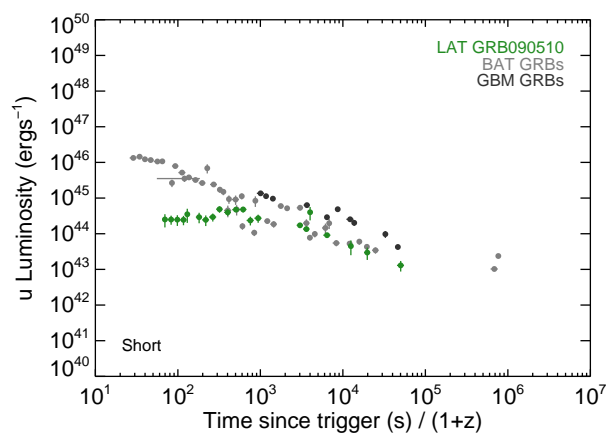}
   \end{center}
   \caption{The u-band normalized afterglow rest-frame luminosity light curves for
    all of the long ({\it top panel}) and short ({\it bottom panel})
    GRBs in our samples.  The BAT, GBM, and LAT GRBs are
    indicated by the different colors.
  \label{fig:olc}}
 \end{figure}

\subsection{UV/Optical} \label{sec:uvot}

The UV/Optical light curves were obtained from the second UVOT GRB
catalog (Roming 
et al., 2011, in-preparation) and combined such that the 5 arcsec
extraction region light curves were used at count rates $>0.5~
\textrm{counts~s}^{-1}$, and 3 arcsec extraction region at lower count
rates.  We combined the UVOT 7 filter light curves (where available
and detected) using the methods of
\cite{oates09} in order to obtain the highest signal-to-noise light
curves.  This involves first normalizing the individual filter light
curves for each GRB to a 
single band, then combining and rebinning.  We chose to normalize all
of the UVOT light curves in this study to the UVOT u-band  
in order to optimize the number of LAT GRB afterglow light curves that could be
used in this comparison, because that was the most commonly used
filter for the LAT burst follow-up.  Note that although 7 of the {\it Swift}
followed-up LAT bursts were detected by UVOT (compared to 8 by XRT),
we could only obtain detailed u-band light curves for five.  The others (GRB
090323 and GRB 100414A) were only detected in the white filter, which
cannot be normalized to the u-band without knowing the shape of the
optical spectrum, and also cannot be easily used to extract
luminosity information because of the flat wide transmission curve of
this filter.  Note that GRB
080916C was detected by XRT, but not UVOT, which is consistent with
the redshift (z=4.35, \citealt{greiner09}).

We convert the UVOT count rate light curves from observed count rate
to flux via the average GRB u-band conversion factor provided by
\cite{poole08}.  We also correct for Galactic extinction, and correct
for host galaxy extinction by fitting broadband (XRT and UVOT) Spectral Energy
Distributions (SEDs) choosing the best fit dust model (MW, LMC, SMC)
for each burst using
the methods of \cite{schady10}.  The 
conversion from flux to luminosity is similar to that of the
X-ray luminosities given in Equation \ref{eq:lx}.  Using the optical
spectral index from the SEDs, we calculate the
u-band k-corrected luminosity as:
\begin{equation}\label{eq:lo}
L_o(t)=4\pi D_L^2 F_o(t)(1+z)^{\alpha_o-\beta_o-1}.
\end{equation}
where $F_o=t^{-\alpha_o}\nu^{-\beta_o}$, and $F_o$ is the u-band flux at time
$t$.

We fit the count rate light curves in a similar manner to that of the
X-ray light curves, except that we allow an additional constant
contribution to the power-law fits to account for the flattening
occasionally observed in UVOT light curves.  This flattening can be due to
either host galaxy contribution or nearby source contamination.  By
simply allowing the fit to include this extra constant, we can
subtract it off and extrapolate the power-law fit to the time of interest.

The u-band rest frame luminosity light curves for the short and long bursts for
each of our BAT, GBM, and LAT samples are shown in Figure
\ref{fig:olc}.  The light curve shapes have not been altered to adjust
for the extra constant.

\subsection{$\gamma$-ray}
Due to the different $\gamma$-ray instruments used to detect the GRBs
in our samples, observational biases cause much of the differences between
these samples.  Likely, the only differences between the GBM and BAT
samples are related to the larger and harder energy range of the GBM,
and the superior sensitivity of the BAT.  However, {\it Swift} has had 6.5
years to collect a sample of GRBs with a wide range of spectral
properties and brightness.  This is reflected in the ranges of the
X-ray and UV/Optical afterglow light curves.  The LAT sample on the
other hand, although a subset of the GBM sample, has inherent
differences.  $50\%$ of the GBM bursts occur within the LAT
field-of-view, but only a small fraction $<5\%$ are detected.  For the
LAT to detect a GRB, the prompt emission must have a very high
fluence.  The factors that may contribute to the high fluence in the 
$30~\textrm{MeV} - 100~\textrm{GeV}$ bandpass include the  
spectrum peaking at relatively high energies, the spectrum having a very
shallow $\beta_{Band}$ index,  and the presence of an additional
hard power law component (as has been detected in several LAT bursts).

In Section \ref{sec:energ}, we discuss burst energetics based upon
observations and limits from afterglow light curves.  The prompt
emission spectral fits used in these calculations were obtained from
several sources.  The BAT spectral fits and fluences come from the
second BAT GRB catalog \citep{sakamoto11}.  
The GBM spectral fits come from either individual burst papers in the
literature, or GCN circulars.  Fluences were recalculated from these
fits in several different bandpasses for the calculation of
$E_{\gamma,iso}$ and prompt emission fluence ratios.

%We extracted GBM spectra for all of the GBM and LAT burst samples over
%the full duration of the bursts, and fit them using {\it rmfit}
%\citep{references}.  Our fits are in approximate agreement with the
%values provided in the GBM team GCN circulars.  We extracted fluences
%in several different energy bands, as will be discussed below, from
%these spectral fits.

%% ANY REANALYSIS OF LAT DATA?  EXTENDED EMISSION LIKELIHOOD DECAYS?

\section{Results}\label{sec:results}
Using our compiled luminosity rest frame light curves and SEDs, we
explored various parameters for differences and similarities between
the BAT, GBM, and LAT burst populations.  The goal was to determine whether
the LAT detected GRBs are fundamentally different from the
normal BAT and GBM samples, or whether they are simply the extreme cases.

Unfortunately, except in the case of the short GRB 090510, we do not
have any early afterglow observations of the LAT bursts, therefore we
limit measurements to times for which all data sets are available.
One day after the trigger in the rest frame is within the BAT,
GBM, and LAT observations, though we also compare some properties at
11 hours (a standard 
observed frame time used in other papers including
\citealt{depasquale06} and \citealt{gehrels08}).  For the 
cases of {\it Swift} bursts, when no data are available at 
this time, we extrapolate from the earlier power-law decay index.

\subsection{Temporal and Spectral Indices}
Using the light curve fits described in Sections \ref{sec:xray} and
\ref{sec:uvot}, we collect the power-law decay indices at a rest frame
time of 1 day.  Typically this is the decay index in the normal
forward shock phase (post-plateau, pre-jet break) in the X-ray light
curves \citep{zhang06,nousek06,racusin09}.  In the cases where the light curves
end prior to 1 day (rest frame), we take the final decay index.  The
UV/optical light curve behavior observed by UVOT 
has a different early morphology, often with an initial rise
followed by a shallow decay and occasionally later
steepening \citep{oates09}. \cite{oates09} observed that the
distribution of $\alpha_o$ after 500 s (post-trigger) is similar to
$\alpha_x$ during the plateau phase, suggesting that the optical
afterglows are also affected by energy injection at early times, or
their temporal profile deviates due to a break (e.g. cooling break) between
the two bands.

\begin{figure}
   \begin{center}
   \includegraphics[scale=0.6,angle=90, bb=100 280 504 684]{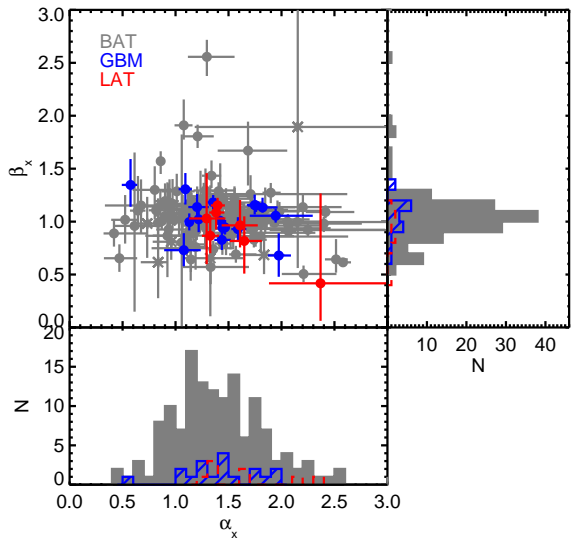}
   \includegraphics[scale=0.6,angle=90, bb=100 280 504 684]{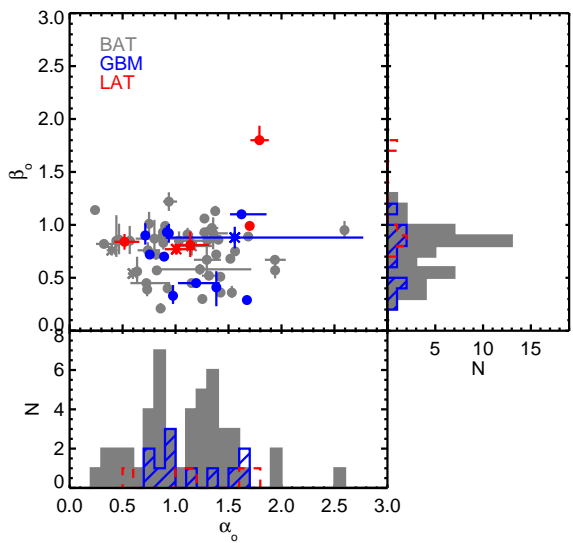}
%% BB 100 280 .. ..
   \end{center}
   \caption{The temporal and spectral decay indices at 1 day in the
     rest frame interpolated from the X-ray ({\it top panel}) and optical
     ({\it bottom panel}) afterglow fits.  The scatter plots and
     histograms show that there are no noticeable differences between
     the BAT, GBM, and LAT populations.
  \label{fig:alpha_beta}}
\end{figure}

Similarly, we also extract the X-ray and optical spectral indices
($\beta_x$ and $\beta_{o}$), and plot them against the temporal
indices in Figure \ref{fig:alpha_beta}.  There is significant scatter
in both the temporal and spectral properties, and no significant
correlation.   

We note that  $\alpha_o$ is systematically lower than
$\alpha_x$ at 1 day in the rest frame, though many of the UVOT light
curves were extrapolated from earlier observations to their expected
later behavior assuming no breaks.

There is little correlation in either the X-ray or optical
temporal and spectral indices, as expected (given the variety of
possible closure relations, \citealt{racusin09}).  From these
distributions, one can see that there are no noticeable differences between the
BAT, GBM, and LAT populations.  Kolmogorov-Smirnov (K-S) tests confirm
that there are no statistically significant differences between the
populations in these measurements (Table \ref{tab:kstest}).

\subsection{Redshift \label{sec:redshift}}
Redshift is another physical quantity that we can evaluate for each of our GRB
populations, with separation into short and long populations.
\cite{jakobsson06} (later updated by \citealt{fynbo09}) showed that
{\it Swift} GRBs are on average at a higher redshift ($\sim 2.2$
versus $\sim 1.5$) than pre-{\it Swift} 
populations, likely due to the superior sensitivity of the {\it
  Swift}-BAT and softer energy range than previous instruments.  It
would therefore follow that the GBM and LAT redshift
distributions may be different.  Of course these are instrumental
selection effects, and as we will discuss in Section \ref{sec:lum},
the LAT bursts tend to have brighter optical afterglows than BAT bursts, therefore
are more likely to have redshift measurements of their optical
transients.  

\begin{figure}
  \begin{center}
    \includegraphics[scale=0.35,angle=90]{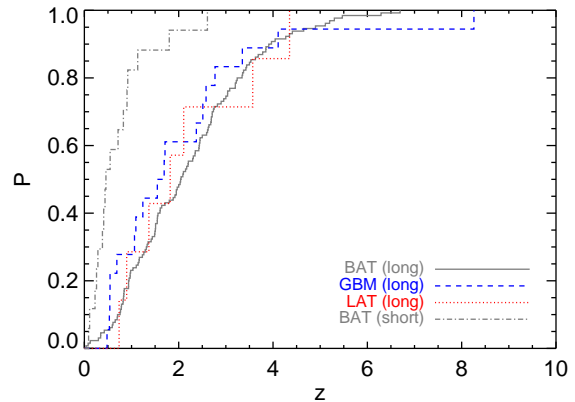}
    \caption{Cumulative distribution functions of the long and short
      burst redshift distributions of the BAT, GBM, and LAT GRB
      populations in our sample.  Note that no distributions are
      plotted for the short GBM and short LAT bursts, because only
      one GRB is present in each of those samples.  \label{fig:zdist}} 
  \end{center}
\end{figure}

With our limited statistics, there are no
significant differences (measured with K-S tests, Table
\ref{tab:kstest}) between the redshift 
distributions of the BAT, GBM, and LAT populations as illustrated in
Figure \ref{fig:zdist}.
However, these are not independent samples, as the GBM bursts were all
also detected by BAT and localized by XRT/UVOT, and these are only the
brightest best localized LAT bursts.  Despite these caveats, there is
no evidence of any differences in redshift distributions. 

\subsection{Environment\label{sec:enviro}}

\begin{figure}
  \begin{center}
  \includegraphics[scale=0.6,angle=90, bb=180 280 550 684]{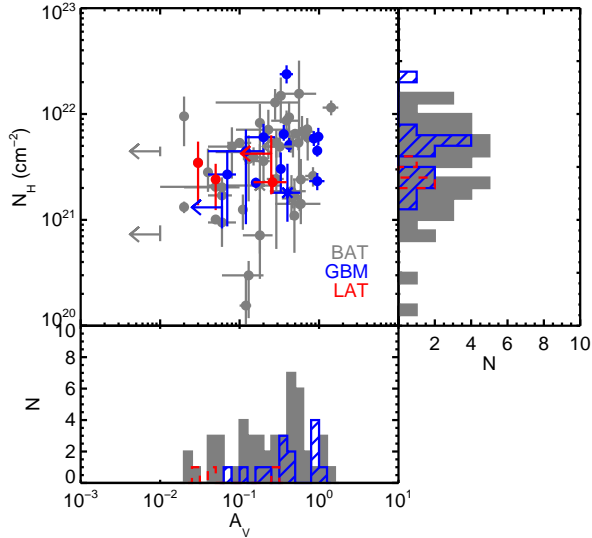}
    \caption{Intrinsic X-ray absorption ($N_H$) plotted against
      visual dust extinction ($A_V$) measured from the SED fits
      described in Section \ref{sec:uvot}.
      \label{fig:av_nh}}
  \end{center}
\end{figure}

Using the parameters from the SED fits,
we can constrain measurements of the X-ray absorption ($N_H$) or
approximate gas content and the optical extinction ($A_V$) or approximate dust
content, in order to learn about the GRB environments.  After we
remove the Galactic absorption and extinction 
contributions, these quantities probe the environment around the GRB 
progenitor and along the line of sight.  Figure \ref{fig:av_nh}
demonstrates these extinction and absorption measurements separated
into the BAT, GBM, and LAT samples.  While a K-S test (Table
\ref{tab:kstest}) does not show any significant differences between the
populations in either $N_H$ or $A_V$, the small number of LAT bursts
for which we could make these measurements tend toward lower values of
$A_V$ with moderate values of $N_H$.  

\begin{figure}
  \begin{center}
  \includegraphics[scale=0.4]{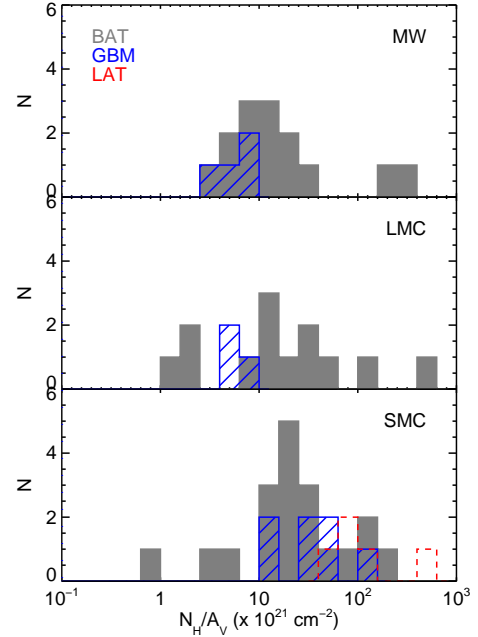}
    \caption{$N_H/A_V$ or gas-to-dust ratio for all of the GRBs in the BAT,
      GBM, and LAT samples for the three best fit dust
      extinctions models (MW, LMC, SMC) for each GRB.
      \label{fig:av_nh_ratios}}
  \end{center}
\end{figure}

The crude gas-to-dust ratios ($N_H/A_V$, Figure \ref{fig:av_nh_ratios}) for
each of the different best fit dust models (MW, LMC, SMC) show two
features that might distinguish the LAT bursts: the LAT bursts are all best
fit by the SMC extinction law, and they tend towards high values of
the gas-to-dust ratio.  Since many of the $A_V$ measurements of the
LAT bursts are upper limits, this makes the corresponding $N_H/A_V$
ratios lower limits, which would only further distinguish the LAT bursts.
%The preference towards the SMC extinction law
%is likely statistical and not physical.  Of the three commonly used
%dust extinction laws, the SMC law is the most simple and featureless
%with the fewest number of parameters and therefore the best fit statistic.
Due to the low $A_V$ values (even $A_V=0$) of the LAT bursts, they can
be fit by any of the the three extinction laws equally well
\citep{schady10}.  The gas-to-dust ratio intrinsically depends on
several factors including the original pre-GRB environmental ratio of
gas-to-dust, and the amount of dust destruction and photoionization during
the GRB event in close proximity to the bursts.  Properties of the GRB
itself such as amount and 
spectrum of the energy output influence the alteration of the environment.
Therefore, understanding differences in the
final gas-to-dust ratio is clouded by these factors which are difficult to
disentangle between properties of the environment and the GRB itself.
This measurement is also model dependent, and can be biased by
assumptions about the spectral model.
Regardless, the LAT bursts appear to have little to no dust along the
line-of-sight compared to the GBM and BAT bursts.  
%The lack of dust
%might indicate either that the progenitors live in younger galaxies
%without significant environmental enrichment, or
%these brighter more energetics GRBs are more apt to destroy the dust
%along the line-of-sight during the GRB.

At this time, there are insufficient statistics on gas and dust
content of the LAT bursts to
draw any strong conclusions from this sample.  Further study with more
objects and broader band data are needed to distinguish any strong
environmental differences between these populations.

\subsection{Afterglow Luminosity}\label{sec:lum}
The luminosity light curves in Figures \ref{fig:xlc} and
\ref{fig:olc} reveal several interesting observational and possibly
intrinsic differences between the BAT, GBM, and LAT populations.  The
GBM and to a higher extent LAT X-ray afterglows are 
clustered much more than the BAT afterglows.  From an instrumental
perspective, BAT is more sensitive to detecting faint GRBs than GBM,
therefore having a wider and fainter distribution of X-ray
afterglows (that correlates with prompt fluence, \citealt{gehrels08}) is
reasonable.  However, due to the correlation between $\gamma$-ray
fluence and X-ray flux, with the LAT GRBs having comparatively extreme fluences
\citep{swenson10,cenko10,mcbreen10,ghisellini10}, we would have
expected the LAT GRBs to be at the 
bright end of the X-ray afterglow distribution.  This distribution
is present in all permutations of these light curves (count rate,
flux, flux density, luminosity), so it is not an effect of one
of our count rate to flux, flux density, or luminosity correction factors.

\begin{figure}
  \begin{center}
    \includegraphics[scale=0.35,angle=90]{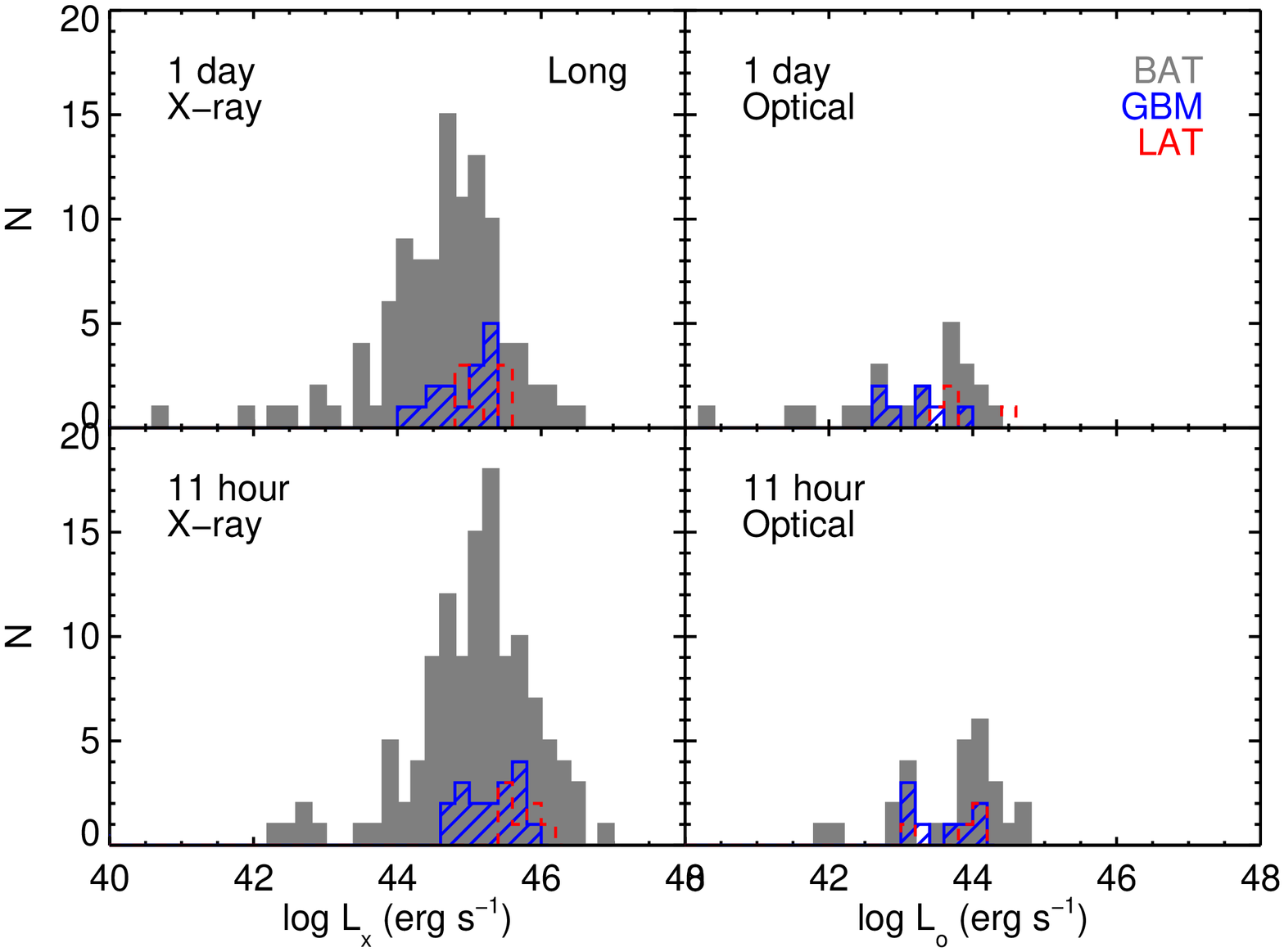}
    \includegraphics[scale=0.35,angle=90]{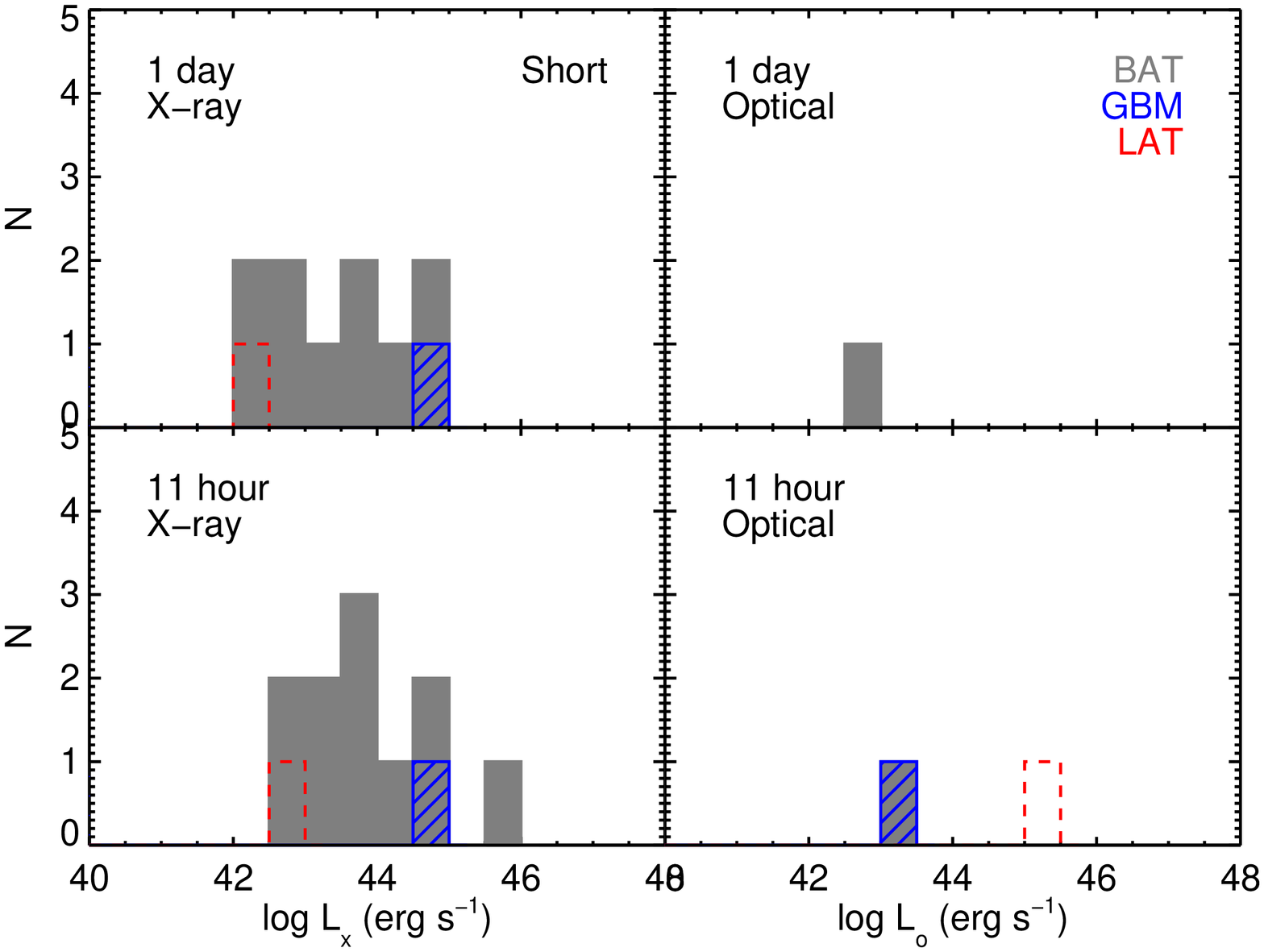}
    \caption{Histograms of X-ray (0.3-10 keV, {\it left column}) and
        optical (u band, {\it right column}) instantaneous luminosity 
      at times of 11 hours ({\it bottom row}) and 1 day ({\it top
        row}) in the rest frame of each GRB, with
      the long bursts ({\it top panel}) and short bursts ({\it bottom panel}) separated.
      Notice that the LAT long burst population luminosities are larger on
      average than that of the other samples, but are not at the
      very bright end of the distribution.
%      {\bf separate long/short???}
      \label{fig:lumhist}}
  \end{center}
\end{figure}

This unexpected distribution is shown more clearly in the histograms
of Figure \ref{fig:lumhist}, demonstrating a cross-section of the
luminosity at 11 hours and 1 day in the rest frame of each GRB.  

\subsection{Energetics} \label{sec:energ}
With this large sample of X-ray and optical afterglows, redshifts, and
simple assumptions about the environment and physical parameters, we
can estimate the total isotropic equivalent $\gamma$-ray energy output
in a systematic way over the same energy range for all of the GRBs in our samples.
Despite the fact that we do not have accurate measurements of
$E_{peak}$ for most of the BAT bursts, we can estimate both $E_{peak}$
(using the power law index correlation from \citealt{sakamoto09}), and either
estimate the Band function or cutoff power law parameters, use typical
values, or use measurements from other 
instruments with larger energy coverage (e.g. Konus-Wind, {\it
  Fermi}-GBM, Suzaku-WAM) especially if they have
constrained $E_{peak}$. 
Using the assumed spectrum for each GRB and the measured redshift, we
integrate over a common rest frame energy range \citep{amati02} of 10
keV to 10 MeV, as:
\begin{equation}\label{eq:eiso}
 E_{\gamma,iso}=\frac{4\pi D_L^2}{(1+z)}\int_{10~keV/(1+z)}^{10~MeV/(1+z)} E~F(E)~dE.
\end{equation}
The functional forms and assumptions are described
in more detail in the appendix of \cite{racusin09}.  Using this
method, we infer a reasonable value of $E_{\gamma,iso}$ for each GRB
in a systematic way.

\begin{figure}
  \begin{center}
    \includegraphics[scale=0.4]{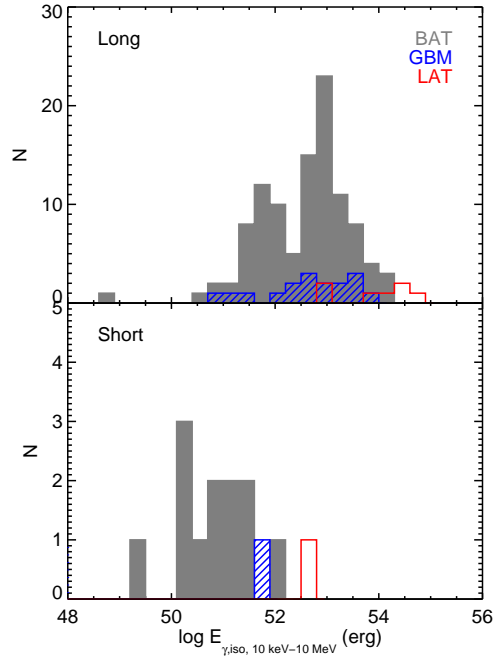}
   \caption{Distribution of $E_{\gamma,iso}$ for both long ({\it top panel})
     and short ({\it bottom panel})
      bursts in the BAT, GBM, and LAT samples.  The GBM and BAT
      distributions are statistically similar.  However, the LAT
      GRBs are on average more energetic than the other samples and
      extend above $10^{55}$ ergs. \label{fig:eiso}}
  \end{center}
\end{figure}

\cite{ghisellini10} and \cite{swenson10} established that LAT GRBs
include some of the most 
energetic GRBs ever detected.  On average, the LAT GRBs have isotropic
equivalent $\gamma$-ray energy outputs ($E_{\gamma,iso}$) that are 1-2
orders of magnitude larger than that of the {\it Swift} bursts (Figure
\ref{fig:eiso}).  Given the
well known correlations between $E_{peak}$ and $E_{\gamma,iso}$
\citep{amati02,amati06}, and the hardness of LAT GRB spectra required
for them to be detected by LAT at all, their large $E_{\gamma,iso}$'s
are not surprising.

This suggests to us that the LAT is preferentially detecting extremely
energetic GRBs compared to previous GRB experiments.  The sensitivity,
large field of view, and large energy range of the LAT make it
especially sensitive to hard bursts.  While the physical origin of the
Amati relation is not well understood, the energetic LAT bursts seem
to qualitatively follow the same relationship.

Applying our characterizations of the optical and X-ray light curves
and SEDs to the energetics, we can infer jet half-opening angles and
collimation-corrected $\gamma$-ray energy outputs ($E_{\gamma}$), or
limits when all observations were either pre- or post-jet break.  Again, the
methods used in these calculations and jet break determination are
described in detail in \cite{racusin09}.

Using the XRT and UVOT data alone, most of the LAT GRB afterglow light
curves (exceptions discussed below) are
best characterized by single power laws, with relatively flat slopes
($\alpha_{o,x}\lesssim 1.8$), with the exception of the poorly sampled GRB
100414A which may have had a break in the large gap between
observations, and the short GRB 090510 which shows an early break to a
steep decay - a behavior suggestive of a ``naked'' short hard burst
\citep{kumar00} that indicates the turnoff of the prompt emission in a
low density environment with either an afterglow too faint to detect
or no afterglow at all.  However, \cite{depasquale10} discussed the possibility
that the break in the optical and X-ray light curves of GRB 090510 at $\sim
2000$ seconds is an early jet break, rather than a naked afterglow
(i.e. steep fall off is either high latitude emission or post-jet break).  The
following calculations use the jet break assumption, but we recommend
caution when examining the energetics of this GRB.

The LAT optical light curves, where sampled well,
show shallow behavior or contamination at late times by the host
galaxy or nearby sources.  This is consistent with the idea that most
of the LAT afterglow observations are pre-jet break (with the
exceptions noted above).  Several recent papers
\citep{mcbreen10,cenko10,swenson10} suggest that when using other
broadband observations (including deep late optical/NIR observations),
some of these bursts do hint at jet breaks, but the 
{\it Swift} data alone are insufficient to constrain jet breaks.  We
will discuss the differences in jet breaks and energetics between this
paper and those of \cite{mcbreen10,cenko10,swenson10} further in Section
\ref{sec:disc}. 

\begin{figure}
  \begin{center}
    \includegraphics[scale=0.32,angle=90]{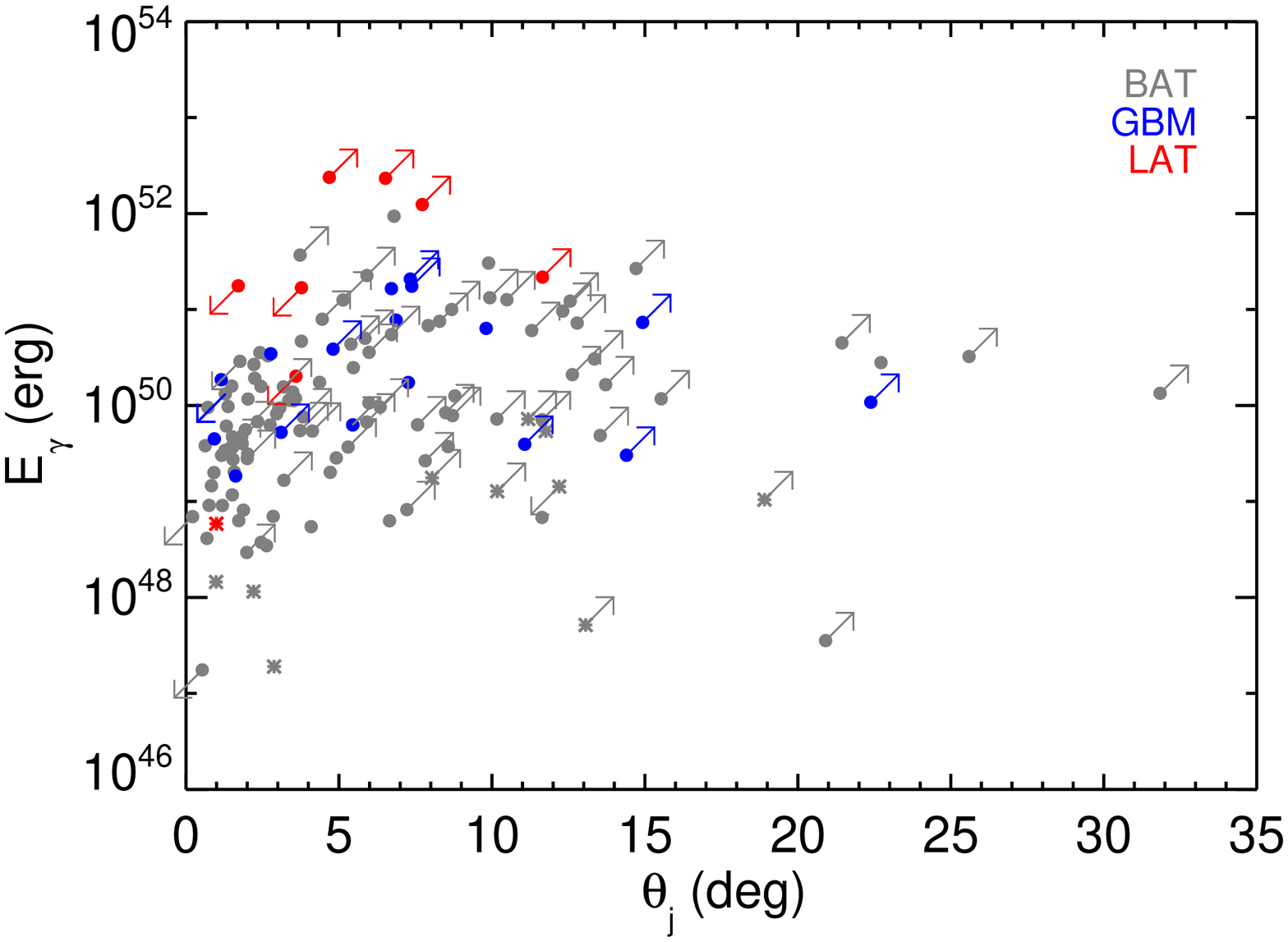}
    \includegraphics[scale=0.3,angle=90]{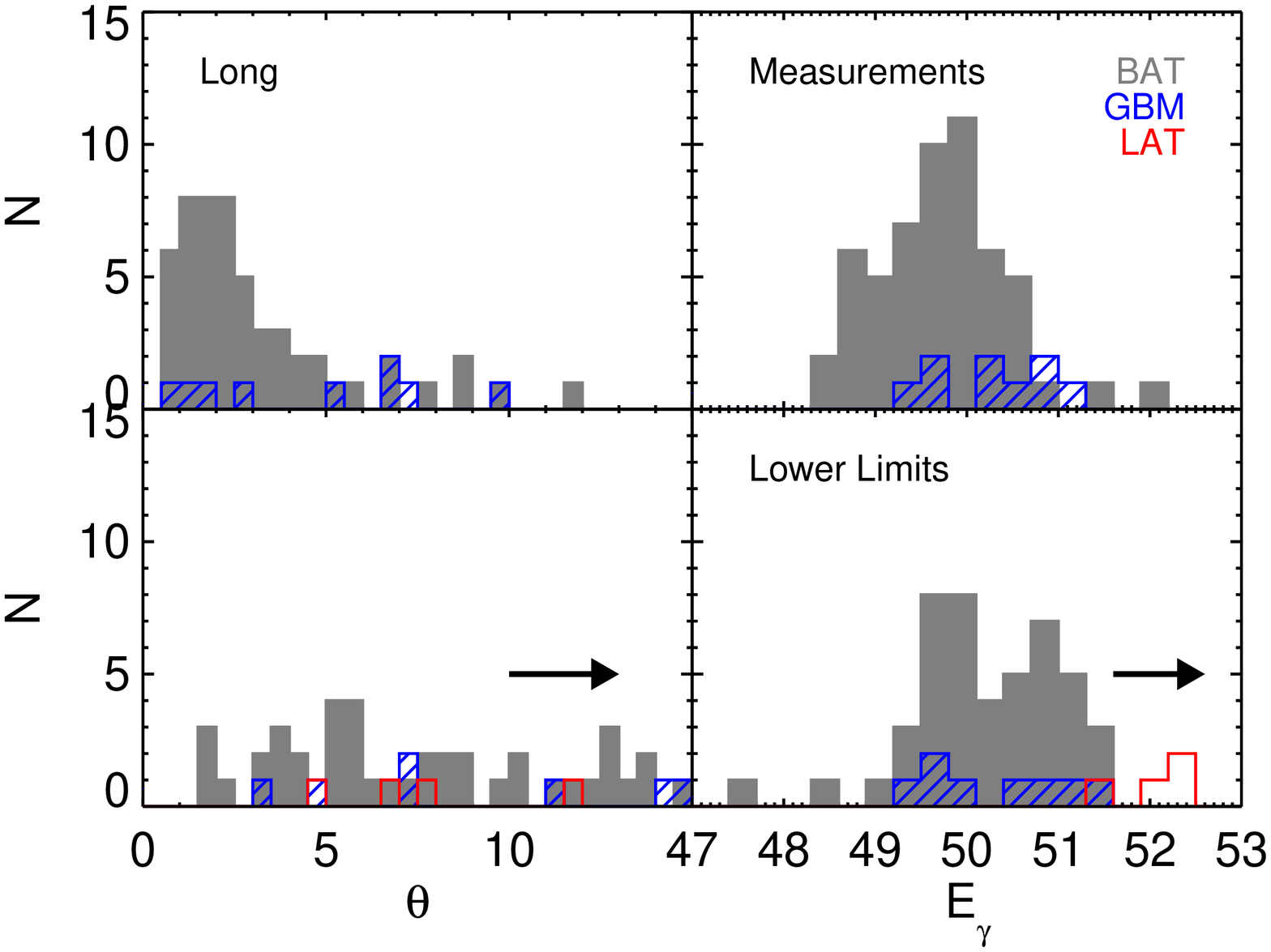}
    \includegraphics[scale=0.3,angle=90]{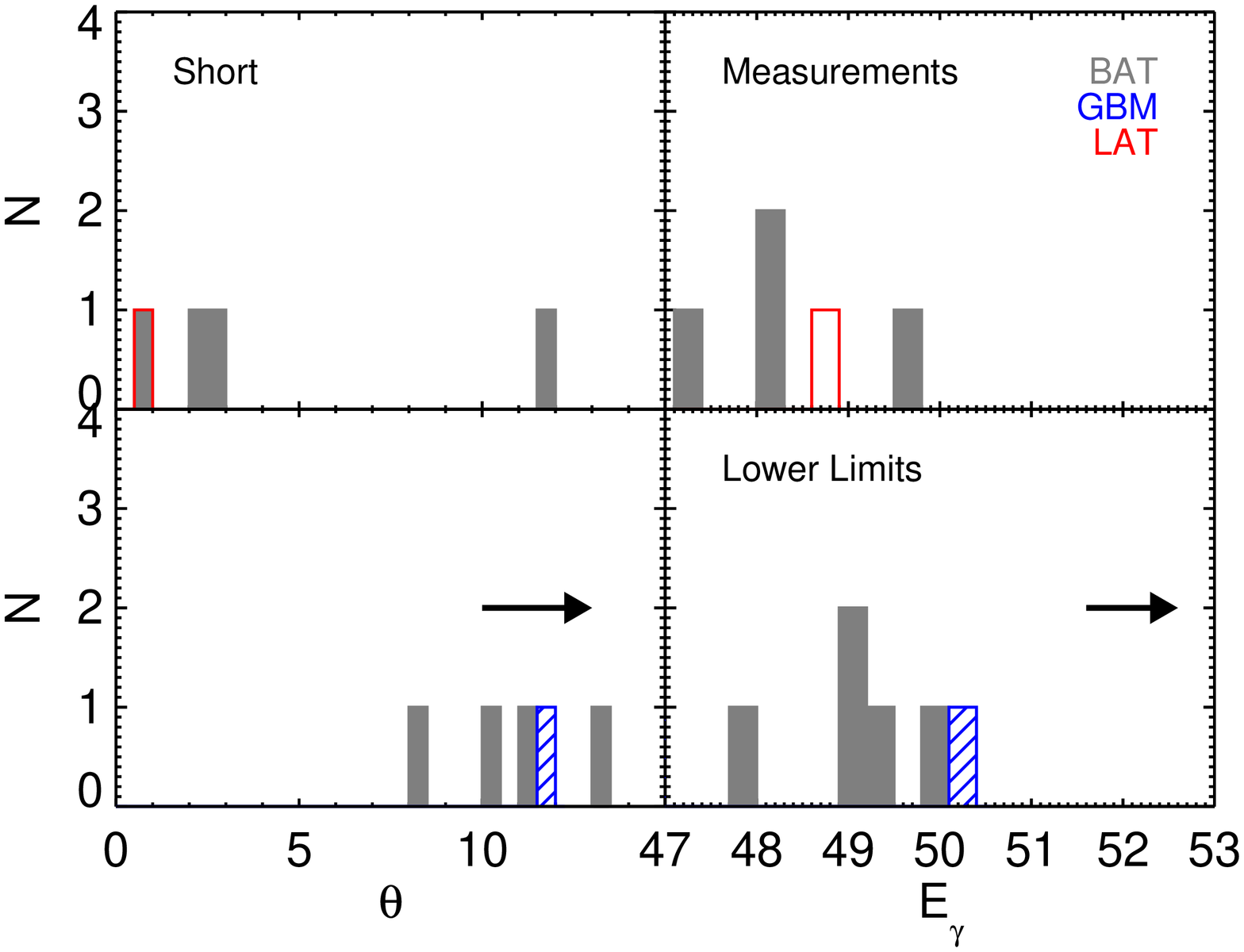}
    \caption{$E_\gamma$ as a function of $\theta_j$ ({\it top panel})
      for the BAT, GBM, and LAT samples, for both long and short
      bursts, including, pre- and 
      post-jet break.  Arrows indicate lower or upper limits on jet
      break times that translate into limits on $\theta_j$ and
      $E_\gamma$.  Histograms of $\theta_j$ and $E_\gamma$ for the
      long bursts ({\it middle panel}) and short bursts ({\it bottom panel})
      including both measurements (upper plots) and upper limits
      (lower plots) as indicated in the top
      panel scatter plot.
      Note that the LAT burst are at the upper end of
      the $E_\gamma$ distribution in comparison to the BAT and GBM
      samples, and are only lower limits.
      \label{fig:energetics}} 
  \end{center}
\end{figure}

If we assume all of the LAT GRBs are pre-jet break (except for GRB
100414A, which may be post jet break, and GRB 090510, which may include
a jet break), and we determine
the presence of jet breaks in the X-ray afterglows of the BAT and GBM
samples using the criteria from \cite{racusin09}, we can evaluate the jet opening
angles and collimation-corrected energetics as a function of these
populations, as shown in Figure \ref{fig:energetics}.

For those GRBs with only lower limits on the jet break times, we use
the time of last detection to determine the lower limit on
$\theta_j$, and therefore also $E_{\gamma}$.  As 
demonstrated in \cite{racusin09}, there are several 
different characteristic times for which one can place limits on jet
breaks, and the large error bars on late-time light curve data points
can mask jet breaks (see also \citealt{curran08}).  However, for Figure
\ref{fig:energetics}, we simply use the time of last detection.

%%% KS-test table

\renewcommand{\thefootnote}{\fnsymbol{footnote}}
\begin{deluxetable}{lccc}
\tablewidth{3in}
\tablecolumns{4}
\tablecaption{K-S Test Probabilities \label{tab:kstest}}
\tablehead{
\colhead{Parameter} & \colhead{BAT-GBM} & \colhead{BAT-LAT} & \colhead{GBM-LAT}}
\startdata
\multicolumn{4}{c}{Long Bursts} \\
\hline
%%new
$\alpha_x$ & 0.78 & 0.14 & 0.54 \\
$\alpha_o$ & 0.93 & 0.44 & 0.63 \\
$\beta_x$ & 0.95 & 0.59 & 0.87 \\
$\beta_o$ & 0.63 & 0.27 & 0.16 \\
$z$ & 0.55 & 0.95 & 0.76 \\
$A_V$ & 0.33 & $9.0\times 10^{-2}$ & $6.8\times 10^{-2}$ \\
$N_H$ & 0.31 & 0.64 & 0.76 \\
$N_H/A_V (\times 10^{21})$ & 0.19 & $5.3\times 10^{-3}$ & $3.0\times 10^{-2}$ \\
$L_{x,11 hr}$\tablenotemark{*} & 0.27 & $2.2\times 10^{-3}$ & $5.3\times 10^{-2}$ \\
$L_{x,1 day}$\tablenotemark{*} & 0.36 & $1.2\times 10^{-2}$ & 0.18 \\
$L_{o,11 hr}$\tablenotemark{*} & 0.38 & 0.29 & $4.8\times 10^{-2}$ \\
$L_{o,1 day}$\tablenotemark{*} & 0.44 & 0.21 & $7.8\times 10^{-2}$ \\
$E_{\gamma,iso}$ & 0.39 & $1.3\times 10^{-3}$ & $1.4 \times 10^{-2}$ \\
$\theta_j$ & 0.15 &  --  &  --  \\
$E_{\gamma}$ & $4.7\times 10^{-2}$ &  --  &  --  \\

\enddata
\tablenotetext{*}{Rest frame time}
\tablecomments{Probabilities of the two distributions being drawn from
the same parent population from K-S tests.  Small values indicate
significant differences between the 
samples. Dashes indicate that there were not enough ($\geq 2$) bursts
that fit relevant criteria to perform a K-S test.  Only long burst
statistics are included here, because there were not enough short hard
bursts to perform K-S tests on the GBM and LAT populations.}
\end{deluxetable}
\renewcommand{\thefootnote}{\arabic{footnote}}
%% should separate out short bursts

\section{Discussion} \label{sec:disc}

We have showed that there are observational differences between the
BAT, GBM, and LAT samples throughout the previous sections.  However,
the difficulty lies in separating out the instrumental selection
effects from the physical differences between GRBs that produce
appreciable $> 100 ~\textrm{MeV}$ emission, and those that do not.

The median and standard deviation of the distributions of the many
observational parameters discussed in the previous and following
sections are presented for each sample in Table \ref{tab:medsig}.

In the following section, we will explore physical explanations for the
observable parameter distributions including a calculation of the
radiative efficiency, and speculate on the origin of the afterglow
luminosity clustering.  We also will compare and contrast the other recent
studies of the broadband observations of the LAT bursts.

\renewcommand{\thefootnote}{\fnsymbol{footnote}}
\begin{deluxetable*}{lcccccc}
\tablewidth{7in}
\tablecolumns{7}
\tablecaption{Parameter Population Characterizations \label{tab:medsig}}
\tablehead{
\colhead{Parameter} & \colhead{BAT long} & \colhead{GBM long} &
\colhead{LAT long} & \colhead{BAT short} & \colhead{GBM short} &
\colhead{LAT short}}
\startdata
$\alpha_x$ & $1.46~(0.56,~130)$ & $1.43~(0.35,~18)$ & $1.58~(0.38,~7)$
& $1.50~(0.82,~12)$ & $1.22$ & $2.19$ \\ 
$\alpha_o$ & $1.21~(0.76,~46)$ & $1.11~(0.35,~10)$ & $1.29~(0.59,~4)$
& $0.72~(0.39,~3)$ & $1.56$ & $1.01$ \\ 
$\beta_x$ & $1.08~(0.37,~130)$ & $1.02~(0.18,~18)$ & $0.90~(0.25,~7)$
& $0.96~(0.34,~12)$ & $1.14$ & $0.79$ \\ 
$\beta_o$ & $0.74~(0.24,~48)$ & $0.75~(0.29,~13)$ & $1.11~(0.47,~4)$ &
$0.70~(0.14,~3)$ & $0.88$ & $0.77$ \\ 
$z$ & $2.21~(1.35,~130)$ & $2.06~(1.88,~18)$ & $2.12~(1.36,~7)$ &
$0.71~(0.65,~17)$ & $1.37$ & $0.90$ \\ 
$A_V~(\textrm{mag})$ & $0.32~(0.28,~44)$ & $0.48~(0.35,~12)$ &
$0.11~(0.13,~3)$ & $0.32~(0.20,~2)$ & $0.40$ & -- \\ 
$log~N_H~(\textrm{cm}^{-2})$ & $20.87~(2.95,~130)$ &
$21.22~(2.53,~18)$ & $19.52~(3.27,~7)$ & $21.17~(0.45,~12)$ & $21.26$
& $20.95$ \\ 
$log~L_{x,11~hr} ~(\textrm{erg~s}^{-1})$ & $45.06~(0.83,~120)$ &
$45.28~(0.39,~17)$ & $45.72~(0.24,~7)$ & $43.26~(2.14,~12)$ & $45.00$
& $42.82$ \\ 
$log~L_{x,1~day}~(\textrm{erg~s}^{-1})$ & $44.62~(0.89,~106)$ &
$44.87~(0.40,~15)$ & $45.19~(0.30,~7)$ & $42.72~(2.47,~11)$ & $44.59$
& $42.08$ \\ 
$log~L_{o,11~hr}~(\textrm{erg~s}^{-1})$ & $43.72~(0.71,~27)$ &
$43.57~(0.47,~8)$ & $44.33~(0.57,~4)$ & $43.12$ & $43.27$ & $43.08$ \\ 
$log~L_{o,1~day}~(\textrm{erg~s}^{-1})$ & $43.11~(0.98,~23)$ &
$43.19~(0.42,~7)$ & $43.86~(0.41,~4)$ & $42.98$ & -- & -- \\ 
$log~E_{\gamma,iso}~(\textrm{erg})$ & $52.56~(0.87,~105)$ &
$52.66~(0.88,~17)$ & $54.01~(0.75,~7)$ & $50.81~(0.74,~12)$ & $51.83$
& $52.59$ \\ 
$\theta_j~(\textrm{deg})$ & $3.53~(3.63,~55)$ & $4.73~(3.20,~9)$ & -- & $4.46~(4.93,~4)$ & -- & $0.99$ \\
\tablenotemark{$\ast$}$\theta_{j,lim}~(\textrm{deg})$ &
$9.19~(6.34,~46)$ & $10.68~(6.36,~8)$ & $7.65~(2.95,~4)$ &
$12.28~(4.13,~5)$ & $11.74$ & -- \\ 
$log~E_{\gamma}~(\textrm{erg})$ & $49.73~(0.71,~55)$ &
$50.30~(0.64,~9)$ & -- & $48.31~(1.03,~4)$ & -- & $48.77$ \\ 
\tablenotemark{$\ast$}$log~E_{\gamma,lim}~(\textrm{erg})$ &
$50.20~(0.80,~46)$ & $50.36~(0.74,~8)$ & $52.04~(0.49,~4)$ &
$48.99~(0.79,~5)$ & $50.15$ & -- \\ 
$log~E_k~(\textrm{erg})$ & $53.41~(0.85,~47)$ & $53.52~(0.43,~12)$ &
$53.69~(0.44,~6)$ & $52.40~(1.09,~3)$ & $53.05$ & -- \\ 
$\eta~(\%)$ & $17.58~(23.34,~47)$ & $19.94~(19.15,~12)$ &
$64.80~(22.53,~6)$ & $6.06~(4.63,~3)$ & $5.69$ & -- \\ 
\enddata
%\footnotetext[\ddag]{$\textrm{erg}~\textrm{s}^{-1}$}
%\footnotetext[\dag]{$\textrm{erg}$}
\footnotetext[$\ast$]{Lower limits on jet opening angles and collimation
  corrected $\gamma$-ray energy output as shown in Figure
  \ref{fig:energetics}}
%\footnotetext[\S]{degrees}
\tablecomments{Mean of the distributions of
  each parameter for the BAT, GBM, and LAT samples, separated
  into long and short bursts. Numbers in parentheses
  indicate standard deviation and the number of objects in each
  parameter distribution.  There is only one object 
  in each of the GBM and LAT short burst samples.}
\end{deluxetable*}
\renewcommand{\thefootnote}{\arabic{footnote}}

\subsection{Radiative Efficiency}\label{sec:eta}
Our first attempt to explore the underlying physics is by calculating
the radiative efficiency of the GRBs at turning their
kinetic energy into radiation during the prompt emission.  We follow the
formulation of \cite{zhang07}, which derives the kinetic energy
($E_k$) from the X-ray afterglow observations, and by comparing the 
$\gamma$-ray prompt emission output, we can estimate a radiative
efficiency: 
\begin{equation}\label{eq:eta}
  \eta=\frac{E_{\gamma,iso}}{E_{\gamma,iso}+E_k}
\end{equation}
where $E_k$ depends on the synchrotron spectral regime
(\citealt{sari98}, $\nu>\nu_c$ or $\nu<\nu_c$) as:
\begin{eqnarray}\label{eq:ek_nu1}
  E_{k,52,\nu>\nu_c}=\left(\frac{\nu F_{\nu}}{5.2\times
    10^{-14}~\textrm{ergs}~\textrm{s}^{-1}~\textrm{cm}^{-2}}\right)^{4/(p+2)}\nonumber\\  
  \times ~D_{28}^{8/(p+2)}(1+z)^{-1}~t_d^{(3p-2)/(p+2)}\nonumber\\
  ~(1+Y)^{4/(p+2)}~f_p^{-4/(p+2)}~\epsilon_{B,-2}^{(2-p)/(p+2)}\nonumber\\
  \times ~\epsilon_{e,-1}^{4(1-p)/(p+2)}~\nu_{18}^{2(p-2)/(p+2)}
\end{eqnarray}
\begin{eqnarray}\label{eq:ek_nu2}
  E_{k,52,\nu<\nu_c}=\left(\frac{\nu F_{\nu}}{6.5\times
    10^{-13}~\textrm{ergs}~\textrm{s}^{-1}~\textrm{cm}^{-2}}\right)^{4/(p+3)}\nonumber\\  
  \times ~D_{28}^{8/(p+3)}(1+z)^{-1}~t_d^{3(p-1)/(p+3)}\nonumber\\ 
  \times ~f_p^{-4/(p+3)}~\epsilon_{B,-2}^{-(p+1)/(p+3)}\nonumber\\
  \times ~\epsilon_{e,-1}^{4(1-p)/(p+3)}~n^{-2/(p+3)}~\nu_{18}^{2(p-3)/(p+2)}.
\end{eqnarray}
All subscripts indicate the convention $X_n=X/10^n$ in cgs units.
$D_{28}$ is the luminosity distance in units of $10^{28}$ cm, $t_d$ is
the time of interest in units of days, $p$ is the electron spectral
index derived from the spectral index $\beta_x$ where:
\begin{equation}\label{eq:beta}
  p=
  \begin{cases}
    2\beta_x+1 & \nu<\nu_c \\
    2\beta_x & \nu>\nu_c \\
  \end{cases}
\end{equation}
and 
\begin{equation}\label{eq:fp}
f_p=6.73\left(\frac{p-2}{p-1}\right)^{(p-1)}(3.3\times 10^{-6})^{(p-2.3)/2}.
\end{equation}

\begin{figure}
  \begin{center}
    \includegraphics[angle=90,scale=0.35, bb=54 54 504 684]{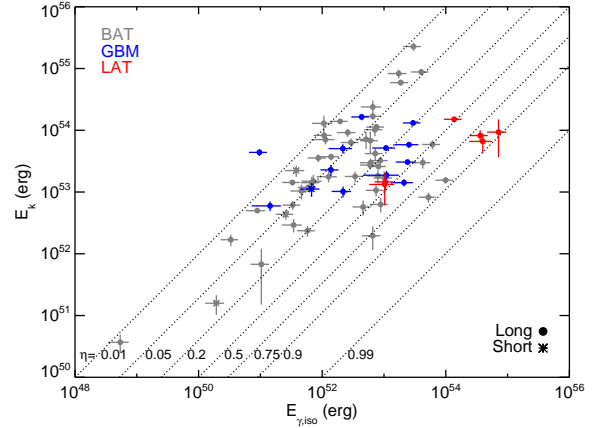}
    \caption{Kinetic energy ($E_k$) derived from the X-ray afterglow
      observations at 11 hours as a function of the prompt emission
      isotropic equivalent $\gamma$-ray energy output
      ($E_{\gamma,iso}$).  The dashed lines indicated different values
      of the radiative efficiency ($\eta$).  The LAT bursts tend to
      have high $E_{\gamma,iso}$ but average $E_k$, and therefore
      higher values of $\eta$ than the BAT or GBM samples.
      \label{fig:efficiency}}
  \end{center}
\end{figure}

We make several simplifying assumptions including using a single typical value for
$\epsilon_e=0.1$, $\epsilon_B=0.01$, and $n=1~\textrm{cm}^{-3}$ as well as ignoring inverse
Compton emission ($Y=1$).  Unlike \cite{zhang07}, we do not determine the
kinetic energy ($E_k$) at the deceleration time or the break time, because we
do not have the temporal context of the shallow decay phase (the
earlier or later canonical phases) in many bursts,
including the LAT bursts.  Instead, we calculated $\eta$ at observed
times of 11 hours and 1 day, but only if there was 
evidence from the light curve morphology and decay slopes of those
measurements being during the normal decay (forward shock) phase
\citep{zhang06,racusin09}.  The efficiency did not change significantly
between these two times, but more GRB X-ray afterglows were in the
normal decay phase at 11 hours than at 1 day.  Therefore, the 11 hour
results are plotted in Figure \ref{fig:efficiency}.  We determined whether a
particular X-ray afterglow was above or below the cooling frequency
($\nu_c$) using the closure relations and assuming the simplest cases
($p>2$, ISM or Wind environments, slow cooling, no energy injection),
similarly to \cite{zhang07}.

Figure \ref{fig:efficiency} demonstrates the relationship between the
kinetic energy and the $\gamma$-ray prompt emission with the range of
radiative efficiencies indicated.
%\begin{equation}\label{eq:hr}
%SR=\frac{S_{\gamma,50-100~\textrm{keV}}}{S_{\gamma,25-50~\textrm{keV}}}
%\end{equation}
%where $S_{\gamma}$ is the fluence in the 50-100 or 25-50 keV bands.
Only 69 GRBs are included in this plot.  The rest of our sample were
excluded either due to not having sufficient information to measure
$E_{\gamma,iso}$, not satisfying the relevant closure relations, lack of a
clear normal forward shock decay, or lack
of emission at time of interest (in this case 11 hours in the observed
frame).  Of the 69 GRBs, 51 are from the BAT sample, 12 from the GBM
sample, and 6 from the LAT sample.

%This figure is similar to Figure 10 in \cite{zhang07}, however, we do
%not differentiate between X-ray Flashes (XRFs) and GRBs.  
Our measured efficiencies for the BAT sample cover a similar range and roughly
agree with their statement from \cite{zhang07}
that given the above assumptions, most ($\sim 57\%$) of BAT bursts
have $\eta<10\%$. 
This statement is not true for the GBM and LAT samples.  In fact, only
$25\%$ of the GBM bursts have $\eta<10\%$, and none of the LAT bursts
have such low radiative efficiencies.  This
suggests that GBM and LAT detected GRBs are on average more efficient
at converting kinetic energy into prompt radiation, which perhaps
explains why they are substantially brighter (higher fluence) than the
BAT bursts with similar afterglow luminosities.  The LAT bursts are at
the high end of the $E_{\gamma,iso}$-$E_k$ %HR-$\eta$
distributions with $\eta>40\%$ for all LAT bursts, and $\eta>80\%$ for
three of the LAT bursts.
%GRB 090902B has the highest $E_{peak}$ and lowest 
%$E_{\gamma,iso}$ of all of the LAT bursts left in this sample (GRB
%090510 has the highest $E_{peak}$ but did not pass the previously
%mentioned cuts), and also has the most luminous X-ray afterglow, and
%second most luminous UV/optical afterglow in the sample.  This
%particular GRB also shows a quasi-blackbody component indicative of a
%fireball photospheric emission (\citealt{ryde10,peer10}, Zhang et
%al. 2010, in preparation).  These atypical properties may
%explain the differences between GRB 090902B and the other LAT bursts
%characterized here, or perhaps the difference is due to a limitation
%of our method and theoretical assumptions.  
These extreme efficiencies may be unphysical and due to a our
assumptions: an internal shock mechanism as is described in the
fireball model \citep{rees98,meszaros02}, a single value of 
$\epsilon_e$ and $\epsilon_B$, no appreciable Compton component, and
a single universal surrounding medium density.

We acknowledge that these differences in the distribution of $\eta$
are degenerate with differences in $\epsilon_e$, $\epsilon_B$, and the
presence of some inverse Compton component.  We also caution that our
estimations of $E_{\gamma,iso}$  for the LAT bursts do not include the
extra spectral power law observed in several of the LAT bursts.  However,
that would make $\eta$ even larger, which perhaps suggests that it is the
internal shock model framework that is not valid.

\subsection{Bulk Lorentz Factor}\label{sec:lorentz}

The bulk Lorentz factor ($\Gamma$) is a fundamental quantity
needed to describe the GRB fireball and therefore interesting to
compare for different 
populations of bursts.  Unfortunately, it is also a difficult
quantity to accurately measure and there are several methods for
placing lower or upper limits on this quantity depending on multiple
assumptions.

The most common technique applied to the {\it Fermi}-LAT detected bursts
\citep{abdo080916c,abdo090510,abdo090902b} 
was originally derived by \cite{lithwick01}, using the highest
energy observed photon to place a lower limit on the $\gamma$-ray pair
production attenuation, setting a lower limit on $\Gamma$.  This
method assumes that the GeV and seed sub-MeV photons are emitted from
the same co-spatial region and are produced by internal shocks.  It 
can produce extreme values of 
$\Gamma\gtrsim 1000$ for the LAT bursts.  \cite{zhao10} and \cite{zou10b} suggest
modifications for this calculation using a two-zone model that assumes
that the sub-MeV and GeV photons are produced at very different radii from the central
engine.  This modification lowers $\Gamma$ to approximately a few
hundred.  Hasco{\"e}t et al. (2011, in-preparation) also demonstrate that when carefully
calculating the pair production attenuation taking into account the
jet geometry and dynamics, $\Gamma$ is reduced by a factor of $\sim
2.5$.

When no high energy (GeV) observations are available, the most common
method to limit $\Gamma$ is to derive it from the deceleration time
of the forward shock which corresponds to the peak time of the optical
\citep{sari99,molinari07,oates09,liang09} or X-ray \citep{liang09} afterglow light
curves.  Often one can only set upper limits on the 
deceleration time because the peak must have occurred prior to the start
of the observations, corresponding to lower limits on $\Gamma$, or be
buried under other components.

There are additional alternative methods to determine $\Gamma$
including putting upper limits on the forward shock contribution to the
keV-MeV prompt emission by looking for deep minima or dips down to the
instrumental threshold between peaks in the
prompt emission light curves \citep{zou10a}.  The typical values of these
upper limits on $\Gamma$ are several hundred.  \cite{peer07} describe
another method that estimate $\Gamma$ from the thermal component
modeling in the prompt emission spectrum using photosphere
modeling. \cite{zhang03} describe yet another 
method to estimate $\Gamma$ using early optical data to constrain
the forward and reverse shock components.  The latter two methods are worth 
further study,  but the application of them to the data presented here
is beyond the scope of this paper.

\begin{figure}
  \begin{center}
    \includegraphics[scale=0.5]{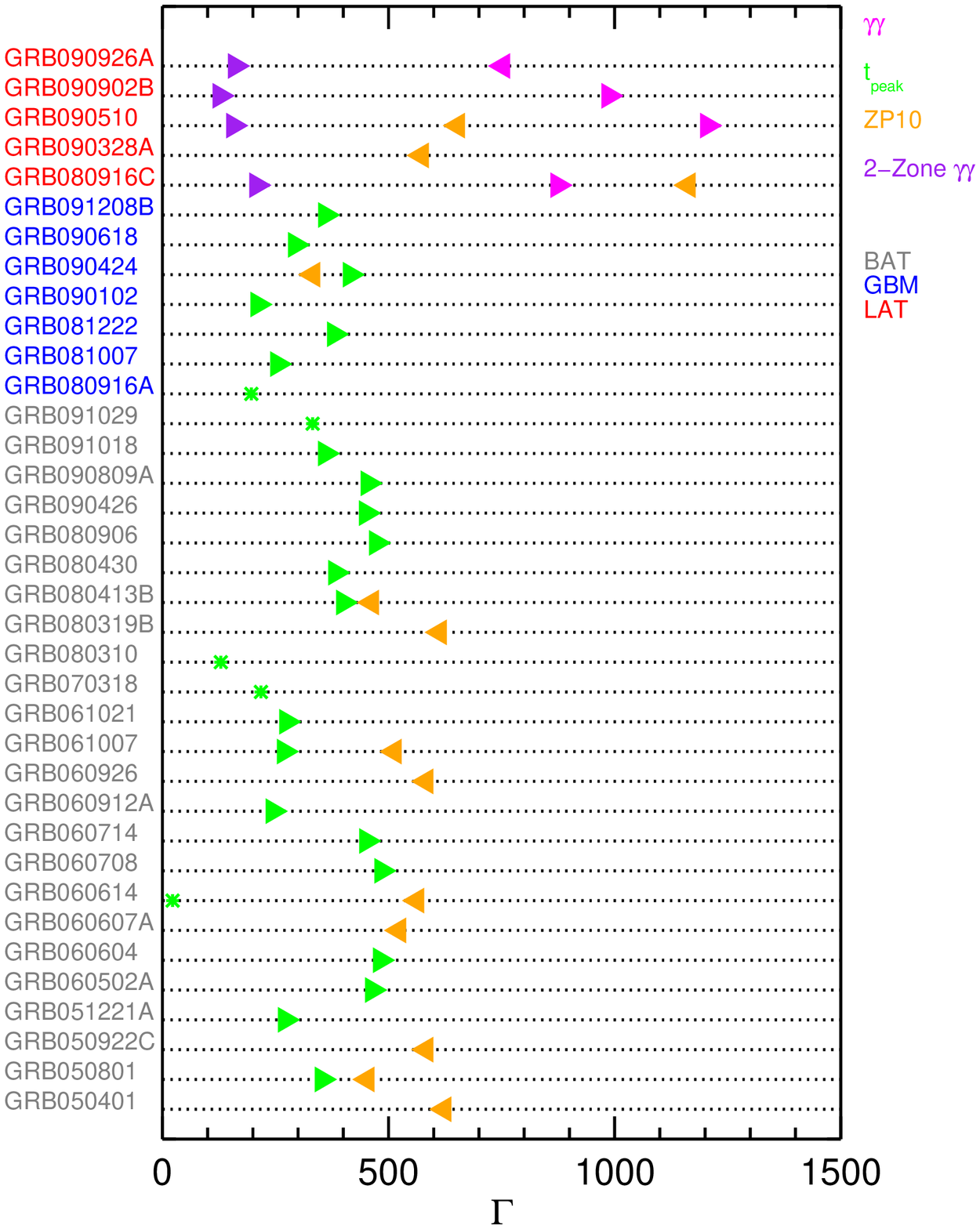}
  \end{center}
  \caption{Estimates of the bulk Lorentz factors ($\Gamma$)
    using the four methods described in Section \ref{sec:lorentz} with
    different colors indicating method.  The four methods are 
    referred to as $\gamma\gamma$ - the pair production attenuation method
    from \cite{lithwick01}; $t_{peak}$ - the forward shock
    estimation from the optical peak described by
    \cite{sari99,molinari07}; ZP10 - the limit on forward shock
    contribution to the sub-MeV prompt emission described in
    \cite{zou10a}; 2-Zone $\gamma\gamma$ - the pair production
    attenuation method assuming the sub-MeV and GeV photon come from
    different sources as described in \cite{zou10b}.  
    A right pointing triangle
    indicates a lower limit, a left pointing triangle an upper limit,
    and an asterisk indicates a measurement.  These measurements or
    limits are either from our own calculations (only
    $t_{peak}$ method) or from the literature.  The bursts for
    which we could make measurements are shown and separated into each
    of the BAT, GBM, and LAT samples.  
   \label{fig:lf}}
\end{figure}

We apply the deceleration time of the optical light curves technique
to our sample for those GRBs with UVOT light curves and measurements of
$E_k$ (derived in Section \ref{sec:eta}).  Unfortunately, due to the
lack of early observations of the LAT bursts, we cannot apply this
method to that sample.  However, we 
collect estimates of $\Gamma$ using the pair-production attenuation
technique from the literature, using the typical one-zone model from
\cite{abdo090902b,abdo080916c,abdo090510,abdo090926a}), as well as the
two-zone estimates from \cite{zou10b}, and compare these limits in Figure \ref{fig:lf}. 

The different methods yield a wide range of $\Gamma$ for each burst
ranging from a few 10s to more than 1000.  However, many of these results
are upper or lower limits.  The assumptions put into each measurement and
method about the geometry and nature of the outflow have a strong
influence on the results.  If we believe that the sub-MeV and GeV photons
are generated in the same co-spatial region, and ignore the two-zone
model estimates, the lower limits on $\Gamma$ of the LAT bursts are
nearly a factor of 2 higher than the BAT and GBM bursts.
Unfortunately, we do not have high 
energy observations of many of the BAT bursts, and we do not have
early optical observations of the LAT bursts, therefore one should compare
measurements of $\Gamma$ for these different samples with caution.

A more careful detailed study of bulk Lorentz factor estimates for the
bursts in this sample would perhaps provide more insight into concrete
differences between the samples.  This would require detailed analysis
of all of the prompt emission light curves and spectra of the bursts
in our sample, and this is beyond the scope of this study.

It is also interesting to note that the high $\Gamma$ limits on
the LAT bursts are reminiscent of a structured jet model, such as the
two-component jet model 
where there is a narrow bright faster core surrounded by a slow wider
jet, with the narrow jet on-axis to the observer
\citep{berger03b,huang04,racusin08}.  \cite{liu10} explore this model
using the broadband data on the LAT GRB 090902B and find an acceptable
fit.  Additional study of the other LAT GRBs in the context of this
model would be needed to draw any stronger conclusions about the sample
as a whole.

\subsection{Afterglow Luminosity Clustering}\label{sec:clustering}
As mentioned in Section \ref{sec:lum}, the X-ray and UV/optical
luminosity distributions of the LAT bursts are narrower than the GBM
and BAT samples.  There are several possible causes or simply related
dependencies, namely, the fact that a larger fraction of the LAT bursts
are in the synchrotron spectral regime $\nu>\nu_c$ ($83\%$) compared
to the BAT and GBM bursts ($50-60\%$), and that the LAT bursts have a
narrow distribution (in log space) of $E_{\gamma,iso}$.  The high radiative
efficiencies of the LAT bursts may be either another cause of the narrow
luminosity distribution or a consequence.  

The region of luminosity parameter space fainter than the LAT bursts
could be limited by the lower detection limits of the LAT instrument, and
the ability to accurately localize only the brightest of the LAT burst
for {\it Swift} follow-up. 
Nearly half of the LAT detections had position errors $>0.5$ degrees
radius, which was simply not practical to initiate follow-up
observations beginning many hours or days after the triggers.  In the
future, if {\it Swift} happens to simultaneously trigger on one of
these fainter long bursts with a marginal LAT detection and a fainter
afterglow, then we will know whether the luminosity clustering is only
limited on the bright end.

%{\bf Need explanation - Look into narrowness of Eiso distribution and
 % $\nu$ compared $\nu_c$ for LAT versus BAT}

\subsection{Luminosity Function}
In addition to the simple redshift distribution comparison, we explore
the luminosity functions of the different populations of GRBs.  We use
the methods of both \cite{virgili09} and \cite{wanderman10}, which
apply different statistical methods to constrain the luminosity
function shapes for the three samples.  Simply due to instrumental
selection effects, the BAT, GBM, and LAT samples probe different
regions of the luminosity function.  GRBs bright enough to trigger
GBM, will be brighter on average than BAT only bursts, because GBM is
less sensitive than BAT.  The LAT GRBs have the highest fluence of the
GBM bursts, and given the similar redshift distributions (Section
\ref{sec:redshift}), are therefore also the most luminous.

The \cite{virgili09,virgili11} method
compares Monte Carlo simulations of the full parameter space drawn
from the full {\it Swift} sample to the specific sample of interest, in order to
be less biased by instrumental selection effects.  Whereas, the
\cite{wanderman10} method does a more tradition fit to the observed
data assuming a single value for the various instrument sensitivity
levels.  Both methods find consistent results for the BAT sample,
fitting to a broken power law with slopes of $0.2$ and $1.4$ for the
less and more luminous ends, respectively, and a break luminosity of
$10^{52.5}~\textrm{erg}~\textrm{s}^{-1}$.  Within the substantial
error bars, the GBM sample resembles the BAT luminosity function, but
does not probe as faint.  The limited LAT sample is even smaller, and
mostly brightward of the break luminosity.  Therefore, it is best fit
by a single power law with slope of 0.3, shallower than the post-break slope of
the BAT and GBM functions.  This shallow slope may suggest some
differences in the parent population of the LAT burst, or may simply
be due to the complicated selection effects of both LAT burst
detection and the subset with accurate enough positions to initiate
follow-up.   Perhaps with a larger LAT sample, this could be studied more
thoroughly.  The bright end of the luminosity function
of the BAT sample may not be well enough probed to accurately
constrain the shape of the luminosity function, and therefore the LAT
bursts are essential tools to study the most luminous GRBs ever detected.

\subsection{Comparisons to Detailed Broadband
  Modeling}\label{sec:others}
%{\bf Add discussion of differences between Cenko, McBreen, Swenson, and us}
Several other recent papers \citep{cenko10,mcbreen10,swenson10} did
detailed broadband modeling of individual LAT bursts.
\cite{swenson10} also made comparisons of the LAT bursts to prompt
emission and afterglow parameters including prompt fluence and
afterglow luminosities, and found that the LAT bursts had higher
X-ray count rates than $80\%$ of BAT bursts at 70 ks.  Note that our
sample is different (only those with redshifts) and likely biased
towards optically brighter bursts, and we measure luminosities rather
than count rates.  However we agree (also with \citealt{mcbreen10}) that the
optical afterglows of the long LAT bursts are in the top half of the
brightness distribution.

We also agree with the aforementioned works, that the LAT bursts
are some of the most energetic GRBs ever observed by any instrument,
and even with collimation corrections or limits on the collimation,
they remain extreme events.  We do not clearly detect any jet breaks
in the LAT bursts using the XRT and UVOT data alone, except for
perhaps the short burst GRB 090510, as discussed in Section
\ref{sec:energ}.  Both \cite{mcbreen10} and \cite{cenko10} claim jet
breaks for some of the LAT bursts, given their ground based deep NIR
and radio 
observations, but most are not well constrained.  Clearly, more late time deep
broadband observations (both space and ground-based) are needed for
the LAT bursts in the future to constrain their total energetics.

\section{Conclusions} \label{sec:conc}

We have systematically characterized the X-ray and UV/optical temporal
and spectral afterglow and prompt emission characteristics,
energetics, and other properties of the GRBs detected only by {\it
  Swift}-BAT, those detected by both {\it Fermi}-GBM and {\it
  Swift}-BAT, and those detected by both {\it Fermi}-GBM and LAT in
order to understand the observational and intrinsic differences
between the burst populations.  There are no significant differences
between the BAT, GBM, and LAT populations in terms of X-ray and
optical temporal power law decays at common rest frame late times, or
spectral power law indices, redshifts, or luminosities.  However, the
distributions of luminosities are much more narrow for the LAT and GBM
samples compared to the BAT sample.  The LAT long burst sample is also
more luminous on average than the BAT sample, but within the same
distribution.  There are significant
differences between the populations in terms of isotropic equivalent
$\gamma$-ray energies ($E_{\gamma,iso}$), prompt emission hardness,
and radiative efficiency.

While in many ways, the late-time ($\sim 1$ day) properties of the LAT
bursts are similar to their lower energy counterparts observed by BAT,
some mechanism fundamentally makes their prompt $\gamma$-ray
production more efficient, or conversely suppresses their afterglows.
As we collect more statistics on this exciting sub-population
of GRBs detected by LAT, we can study luminous and energetic extremes.
Studying the 
afterglows of the LAT burst, especially at early times, will also help
us to understand the additional components (extra spectral power-law
and extended $>100$ MeV emission) observed in many of these LAT
bursts.  Additional future simultaneous triggers between BAT and LAT will
provide more information on the early broadband behavior of LAT
bursts.

\acknowledgements
We gratefully acknowledge A. Fruchter, A. Pe'er, K. Misra, and
J. Graham for helpful discussions.  We also thank the anonymous
referee for their detailed comments.

\bibliographystyle{apj}
\bibliography{bib}

\end{document}